\documentclass{osa-article}
\usepackage{float}
\usepackage{cite}
\usepackage{lineno}
\usepackage{graphicx}
\usepackage{amsmath}

\journal{oe}

\begin{document}

\title{Few-cycle vortex beam generated from self-compression of mid-infrared femtosecond vortex beam in thin plates}

\author{Litong Xu\authormark{1}, Dongwei Li\authormark{2}, Junwei Chang\authormark{2}, Tingting Xi\authormark{1,3} and Zuoqiang Hao\authormark{2,4} }

\address{\authormark{1}School of Physical Sciences, University of Chinese Academy of Sciences, Beijing 100049, China\\

\authormark{2}Shandong Provincial Engineering and Technical Center of Light Manipulations \& Shandong Provincial Key Laboratory of Optics and Photonic Device, School of Physics and Electronics, Shandong Normal University, Jinan 250358, China\\
}

\email{\authormark{3} ttxi@ucas.ac.cn \authormark{4}zqhao@sdnu.edu.cn } 


\begin{abstract}
We demonstrate theoretically that few-cycle vortex beam with subterawatt peak power can be generated by self-compression of mid-infrared femtosecond vortex beam using the thin-plate scheme.
The 3 $\mu$m femtosecond vortex beam with input duration of 90 fs is compressed to 15.1 fs with the vortex characteristics preserved.
The conversion efficiency is as high as 91.5\% and the peak power reaches  0.18 TW.
The generation of the high-peak-power few-cycle vortex beam is owing to the proper spatiotemporal match by this novel scheme, where the spectrum is broadened enough, the negative group velocity dispersion can compensate the positive chirp induced by nonlinear effects, and multiple filamentation is inhibited for the keeping of the vortex characteristics.
Our work will help to generate isolated attosecond vortices, opening a new perspective in ultrafast science.
\end{abstract}

\section{Introduction}

Vortex beams have been widely studied in the last few decades, for both the intriguing physics involved and potential applications \cite{shen2019optical}.
The orbital angular momentum (OAM) induced by the helical phase wavefront connects the classical and quantum physics, offering a convenient tool to study quantum information and quantum computation \cite{fickler2016quantum}.
It also provides a new freedom to influence the light matter interactions, facilitating atom trapping and guiding \cite{franke2008advances}. 
When the OAM is introduced to ultrashort pulses, more fascinating phenomena arise in the nonlinear propagation, such as the spatiotemporal vortices \cite{jhajj2016spatiotemporal}, vortex algebra \cite{hansinger2014vortex} and orbital-to-spin angular momentum conversion \cite{fang2021photoelectronic}.
One of the most attractive topics concerning femtosecond vortex beams is the high harmonic generation (HHG), which is an effective tool to generate extreme-ultraviolet vortex beams \cite{rego2019generation} and trains of attosecond light vortices \cite{geneaux2016synthesis}.
The attosecond light vortices can provide unique light sources for the fundamental study and diagnosis tool in ultrafast science \cite{cireasa2015probing}.
However, there is no report on the generation of isolated attosecond light vortices due to the lack of few-cycle vortex beams \cite{sansone2006isolated}. 

Few-cycle vortex beams are hard to be realized by applying phase encoding elements to few-cycle Gaussian beams due to chromatic aberrations.
They may be obtained by post-compression of supercontinuum vortices which have been realized by the four-wave frequency mixing approach \cite{hansinger2014vortex} and the thin-plate scheme \cite{Xu22}.
However, the post-compression of these supercontinuum vortices will cause inevitable energy loss \cite{seo2020high}.
As an alternative way, self-compression is of particular interest for its convenience to generate few-cycle pulses with high conversion efficiency \cite{wagner2004self,nagar2021study}.
Although it is well-known that self-compression may occur when intense femtosecond pulses propagate in media with anomalous dispersion, self-compression of femtosecond vortex beams has not been studied.
This is because the strong nonlinear effects and initial noise usually lead to multiple filamentation, which will destroy the vortex characteristics of the initial beam \cite{neshev2010supercontinuum}.
To solve this problem and generate high power few-cycle vortex beams, the thin-plate scheme is a promising candidate.
The thin-plate setup consists of several fused silica plates which can divide the nonlinear propagation into several parts and inhibits destructive interference of multiple filaments \cite{lu2014generation}.
By using this setup, we have obtained powerful supercontinuum vortices in our previous work \cite{Xu22}.
It has been demonstrated that the phase profile of incident laser can be preserved after the spectral broadening process occurred in the thin plates.
On the other hand, by using the thin-plate setup, mid-infrared (MIR) femtosecond laser pulses can be self-compressed to few-cycle pulses \cite{qian2021few}.
In the fused silica plates, the strong negative group velocity dispersion (GVD) can compensate the positive chirp induced by self-phase modulation (SPM). 
Therefore, the thin-plate scheme should have a very high feasibility of generating high power few-cycle vortex beams.

In this paper, we simulate the self-compression of MIR femtosecond vortex beam with central wavelength of 3 $\mu$m in the thin-plate setup.
90 fs initial femtosecond vortex pulse is successfully compressed to 15.1 fs pulse preserving the initial vortex phase distribution.
The soliton-like self-compression by using the scheme leads to a very high conversion efficiency and compression ratio.
The results demonstrate that this novel scheme satisfies the complex spatiotemporal requirements for generating high-peak-power few-cycle vortex beam, where spectrum broadening, dispersion compensation, and inhibition of multiple filamentation can be realized simultaneously.
The generation of the high power few-cycle vortex beam will help to generate isolated attosecond vortices, opening a new perspective in ultrafast science.

\section{Simulation}

The propagation of MIR femtosecond vortex beam in the thin-plate scheme can be described by the nonlinear envelope equation coupled with the electron density equation, which can be written as \cite{berge2009self}:

\begin{equation}
 \begin{aligned}
\frac{\partial{E}}{\partial z}=&\frac{i}{2 k_{0}} T^{-1} \nabla_{\perp} E+\frac{i\omega_{0}}{c} n_{2} T \int_{-\infty}^{t} \mathcal{R}\left(t-t^{\prime}\right)\left|E\left(t^{\prime}\right)\right|^{2} d t^{\prime} E\\
&+i \widehat{D} E-\frac{ik_{0}}{2 n_{0} \rho_{c}} T^{-1} \rho E-\frac{U_i W(|E|)}{2|E|^2}(\rho_{nt}-\rho)E
\end{aligned}
\label{E}
\end{equation}
\begin{equation}
 \frac{\partial \rho}{\partial t}=W(|E|)(\rho_{nt}-\rho)-\frac{\rho}{\tau_{r e c}},
 \label{ne}
\end{equation}
where $T=1+\frac{i}{\omega_{0}} \partial_{t}$, $R(t)=(1-f) \delta(t)+f \Theta(t) \frac{1+\omega_{R}^{2} \tau_{R}^{2}}{\omega_{R} \tau_{R}^{2}} e^{-t / \tau_{R}} \sin \left(\omega_{R} t\right)$, $f=0.18$ for silica and 0.5 for air, $\widehat{D}=\sum_{n \geq 2}\left(\frac{k^{(n)}}{n !}\right)\left(i \partial_{t}\right)^{n}$, $k^{(n)}=\partial^{n} k /\left.\partial \omega^{n}\right|_{\omega_{0}}$.
$k_0$ is the wavenumber corresponding to central wavelength of $\lambda_0=3$ $\mu$m.
The nonlinear refractive index $n_2$ for fused silica is taken to be $2\times 10^{-16}$ cm$^2$/W, according to recent experiment \cite{patwardhan2021nonlinear}.
The dispersion relations of fused silica and air are taken from \cite{tan1998determination,mathar2007refractive}, applicable to simulated MIR range.
$\rho_{c} =1.2\times 10^{20} \ \mathrm{cm}^{-3}$ is the critical plasma density. The ionization rate $W(|E|)$ for air is calculated from the PPT model (Perelomov, Popov, and Terentev) \cite{perelomov1966ionization,berge2007ultrashort}, and for fused silica we use the Keldysh rate \cite{keldysh1965ionization}.
Other parameters are the same as those used for 800 nm \cite{Xu22}.

The initial field is a singly charged Lagarre-Gaussian beam:
\begin{equation}
E(r,\varphi,t,z=0) = E_0(r/w)e^{-r^2/2w^2}e^{i\varphi}e^{-t^2/0.72\tau^2}e^{-ik_0r^2/2f},
\end{equation}
where the beam radius $w=200\ \mu$m, pulse duration (FWHM) $\tau=$ 90 fs, $f=0.3$ m, $\varphi$ is the azimuthal angle and the pulse energy is 3 mJ. 
10\% random amplitude perturbation is introduced to the initial field.

To obtain few-cycle pulses, several conditions should be satisfied.
First, the negative GVD of plates is high enough to compensate the positive chirp induced by SPM, and we find a central wavelength of 3 $\mu$m meets the requirement. 
We also tried 2 $\mu$m, but in that case pulse splitting occurs, making the chirp more complicated, and only sub-3-cycle pulses can be acquired.
Second, the intensity in each plate is within a proper range.
On the one hand, it is high enough to trigger supercontinuum generation.
On the other hand, the intensity is moderate so that the beam can propagate a longer distance in fused silica before beam collapse, which facilitates pulse compression.
In practice we find 10 -- 20 TW/cm$^2$ a suitable range, which can also avoid damage to fused silica \cite{PhysRevLett.101.213901}.
Third, the vortex ring keeps a relatively uniform intensity distribution, so that multiple filaments are inhibited and the vortex phase can be transferred to newly generated spectral components.
Based on the above considerations, five 300-$\mu$m-thick fused silica plates are used in our simulation, and their spacings vary from 20 mm to 60 mm.
An optimized configuration is shown in Fig. \ref{fig1}.

\section{Results and discussion}

\begin{figure}[htbp]
\centering

	\includegraphics[width=0.95\linewidth]
	{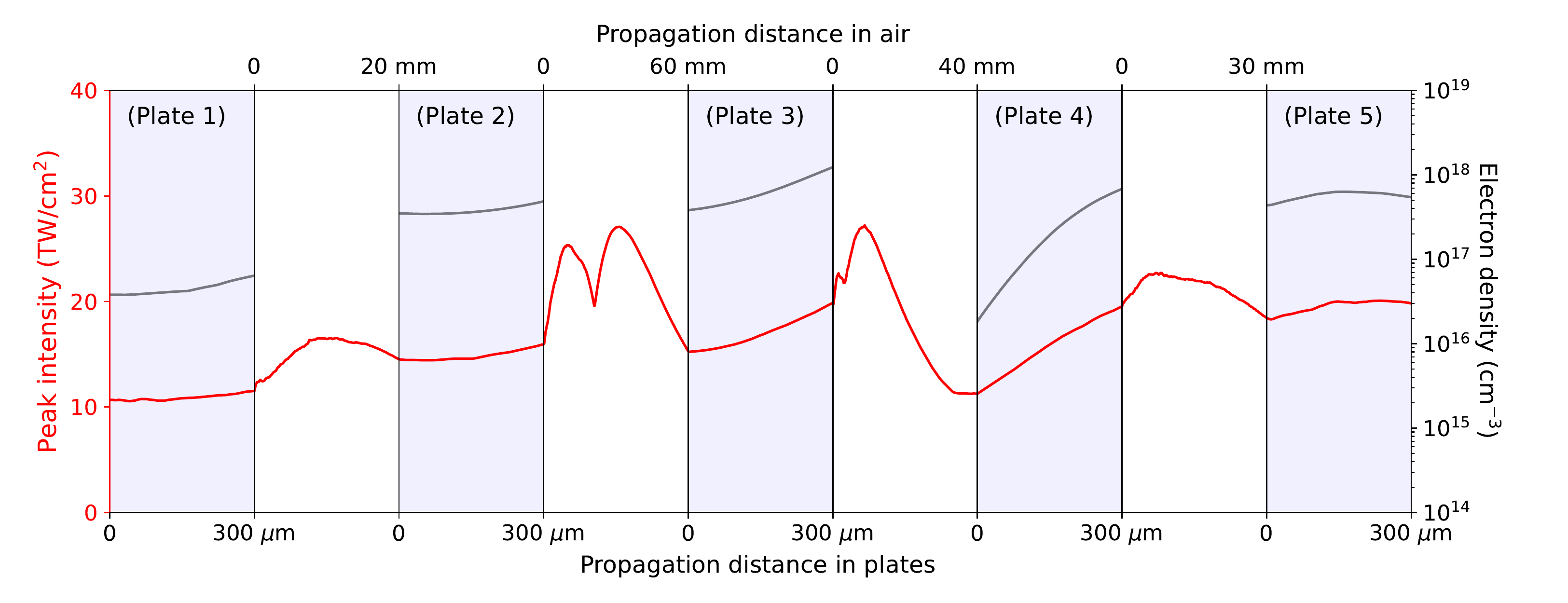}

\caption{Evolution of peak intensity and electron density.}
\label{fig1}
\end{figure}

First, we take an overall view of the propagation process.
The evolution curves of peak intensity and electron density are shown in Fig. \ref{fig1}.
Although there is noticeable increase of peak intensity in each plate, the intensity varies between 10 and 20 TW/cm$^2$, which will not damage the plate but is high enough to trigger spectral broadening.
The obvious increase of intensity in each plate is caused by the temporal compression of the pulse, as will be discussed further below.
After each plate, the vortex beam self-focuses in air, leading to the rapid increase of intensity.
This phenomenon has been investigated in detail for a Gaussian beam, where the nonlinear phase accumulated in the fused silica acts as an effective lens, making the beam refocus in air \cite{berge2009self}.
During the whole propagation process, the electron density is kept in a relatively low level in plates ($< 2 \times 10^{18}$ cm$^{-3}$), and is negligible ($< 10^{14}$ cm$^{-3}$) in air.
The range of the intensity and electron density also ensure that multi-filamentation is inhibited and the intensity distribution of the vortex ring is relatively uniform, so that the characteristics of vortex will be kept during pulse compression.

\begin{figure}[htbp]
\centering

	\includegraphics[width=0.8\linewidth]
	{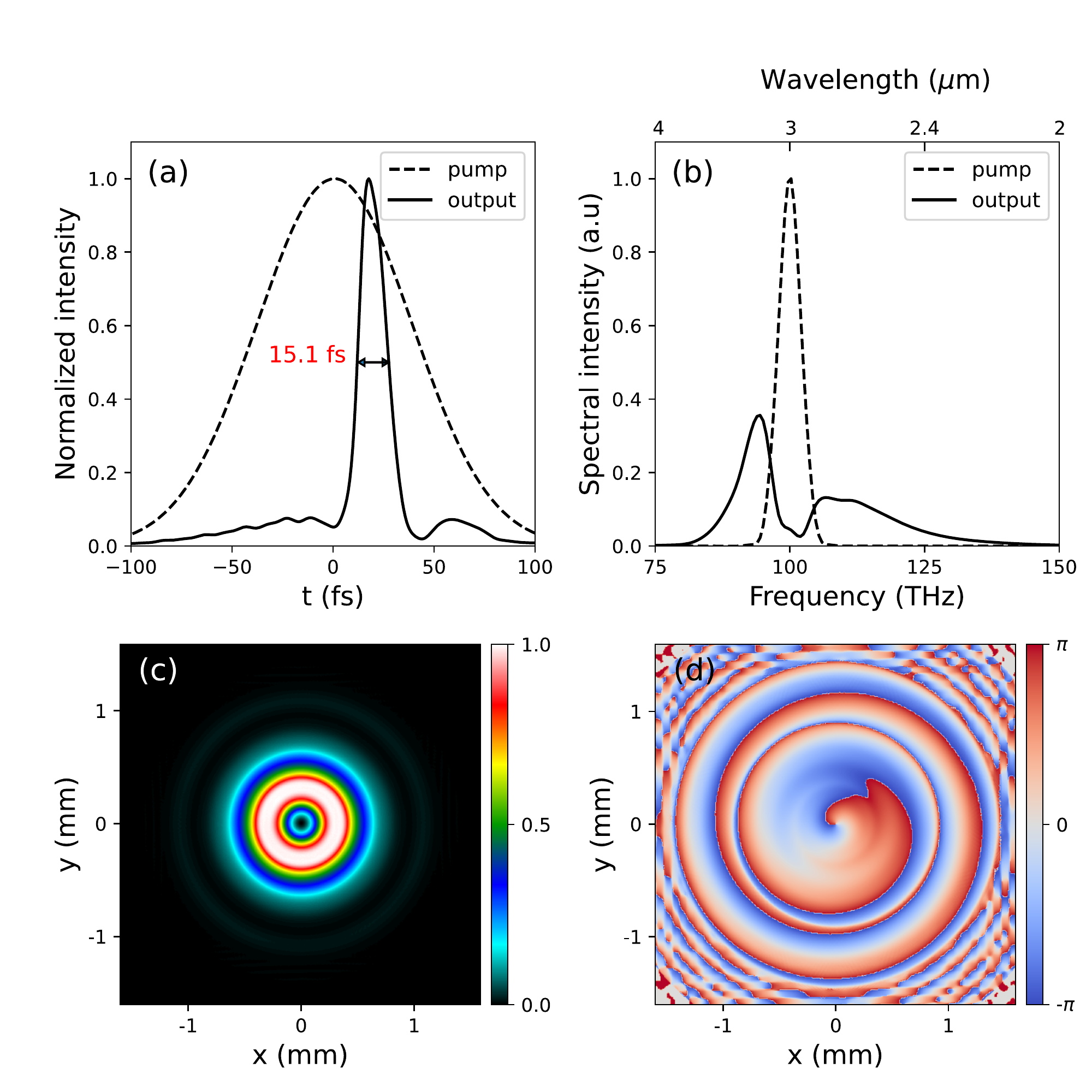}
\caption{Spatiotemporal characteristics of MIR laser pulse after Plate 5: (a) temporal profile, (b) spectral intensity, (c) transverse fluence distribution, (d) transverse phase distribution at temporal intensity maximum.}
\label{fig2}
\end{figure}

After propagation in five plates, the pulse is compressed to 15.1 fs which is about 1.5 optical cycles, with the spectrum spanning from 2 $\mu$m to 4 $\mu$m, as shown in Figs. \ref{fig2} (a) and (b).
Moreover, this few-cycle pulse keeps the ring-shaped intensity distribution with a central singularity in the transverse plane, as shown in Fig. \ref{fig2} (c).
The spiral phase profile in Fig. \ref{fig2} (d) indicates that this few-cycle pulse keeps the same topological charge as the initial laser pulse.
These results suggest that we obtain few-cycle vortex beam via self-compression of MIR femtosecond vortex beam in the thin plates.
Due to the low ionization loss in the scheme, this few-cycle vortex beam has an output energy of 2.75 mJ, corresponding to peak power of 0.18 TW and a high conversion efficiency of 91.5\%.

\begin{figure}[htbp]
\centering
	\includegraphics[width=\linewidth]
	{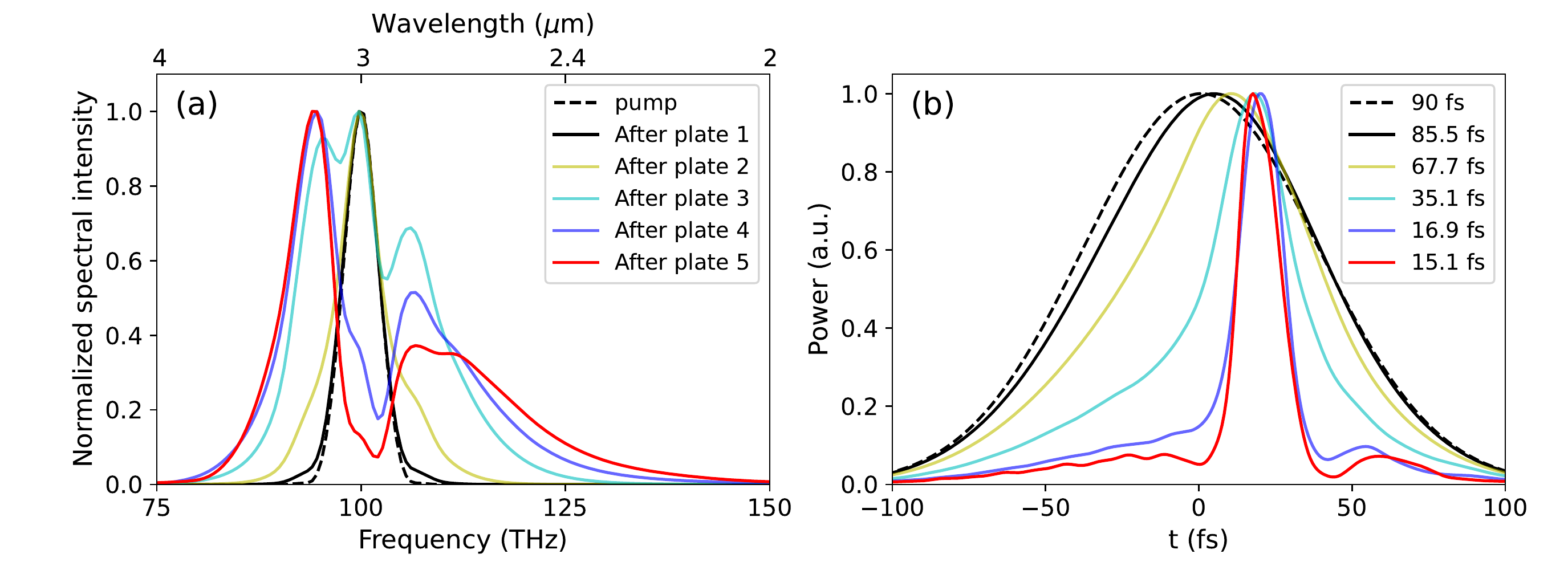}
\caption{(a) Spectra and (b) temporal power distribution of the laser pulse after each plate.}
\label{fig3}
\end{figure}

To investigate the generation mechanism of the few-cycle vortex beam, we plot the spectra and pulse profiles after each plate, as shown in Fig. \ref{fig3}.
At first, we see a symmetric broadening of spectrum, indicating that SPM is predominant in the first plate.
Starting from Plate 2, the spectrum extends to the blue side obviously.
This is because the frequency shift induced by SPM and ionization that can be described by $\Delta \omega / \Delta z \propto \partial_{t} \rho(r, t) / 2 n_{0} \rho_{c}-n_{2} \partial_{t} I(r, t)$, where the first term has the same magnitude as the second term.
Although the electron density is not high, the small value of $\rho_c$ makes SPM and ionization both contribute greatly to spectrum broadening.
After five plates, an octave-spanning spectrum that spans from 2 $\mu$m to 4 $\mu$m is obtained.
The evolution of the pulse profiles in Fig. \ref{fig3}(b) shows that the pulse undergoes remarkable compression in the plates due to strong negative GVD.
Note that during the compression, no obvious steep tailing edge is observed, for which quantitative analysis is given below.
The Kerr effect makes the refractive index of pulse peak higher than that of the pulse leading edge, given by $\Delta n=n_\text{peak}-n_\text{lead}=n_2\Delta I$, which shifts the pulse peak towards the tailing edge at $t>0$.
The relative delay time per meter of the pulse peak and leading edge is thus given by $\Delta t_\text{Kerr}=1/v_\text{peak}-1/v_\text{lead}=\Delta n/c$.
Similarly, the GVD induced delay time of the pulse peak and leading edge is estimated to be $\Delta t_\text{GVD}= 2\pi k^{(2)}\Delta f$, where $k^{(2)}= -534$ fs$^2$/mm is the GVD coefficient of fused silica at 3 $\mu$m, and $\Delta f$ is the central frequency difference of the pulse peak and leading edge.
Assuming $\Delta I=10$ TW/cm$^2$ and $\Delta f=20$ THz, we get $\Delta t_\text{Kerr}=6.7$ ps and $\Delta t_\text{GVD}=-67$ ps.
Since $|\Delta t_\text{GVD}|$ is much larger than $|\Delta t_\text{Kerr}|$, self-compression induced by GVD is more significant than self-steepening induced by SPM.
As the pulse tailing edge becomes steep, the pulse leading edge is also compressed rapidly towards the tailing edge.
Consequently, both the leading and tailing edges become steep in the output few-cycle pulse, thus the traditional self-steepening effect is not obvious here.
Meanwhile, this leads to soliton-like self-compression without pulse splitting, which enables a higher compression ratio.

\begin{figure}[htbp]
\centering
	\includegraphics[width=\linewidth]
	{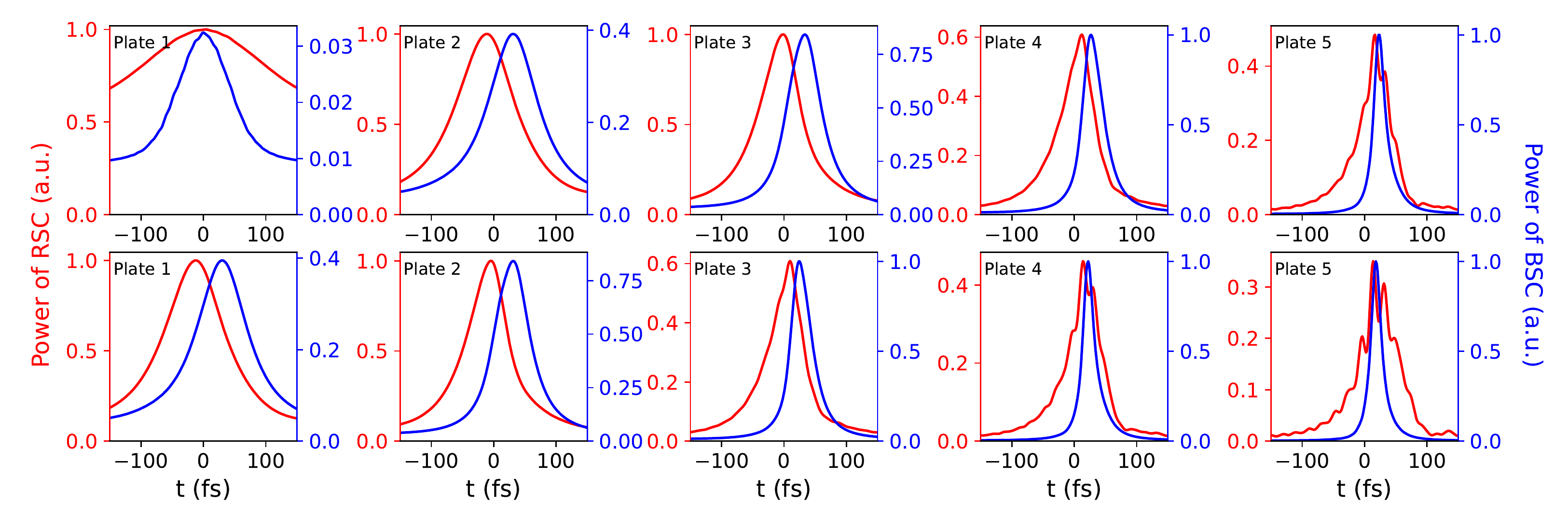}
\caption{Power distribution of the redside component (RSC, $>3.2\ \mu$m) and blueside component (BSC, $<2.8\ \mu$m) in the temporal region for the front surface (top row) and back surface (bottom row) of each plate.}
\label{fig4}
\end{figure}

In Fig. \ref{fig4} we give a dynamic view of how the chirp is compensated via the GVD of plates.
We select two spectral regions, redside component (RSC, $>3.2\ \mu$m) and blueside component (BSC, $<2.8\ \mu$m).
For each spectral component, we filter it and perform inverse Fourier transform to get its wave packet in the time domain.
Figure \ref{fig4} shows the temporal distribution of the two components at the front surface and back surface of each plate.
Due to SPM,  the RSC and BSC are located symmetrically at the back surface of the first plate, with RSC in the pulse leading edge and BSC in the tailing edge.
Although ionization also contributes greatly to the spectral broadening in the following plates, the temporal distribution of the BSC is not distorted obviously, and only the proportion of the BSC increases.
With further propagation in plates and  broadening of the spectrum (Fig. \ref{fig3} (a)), the effect of GVD becomes more significant.
It can be seen that, the RSC and BSC continuously draw close (bottom row in Fig. \ref{fig4}), which leads to the compression of pulse.
Besides, after Plate 3 the duration of BSC becomes much shorter than RSC.
This is in accordance with the extra spectral blueshift induced by ionization shown in Fig. \ref{fig3}.
Finally, after Plate 5, RSC and BSC are synchronized in time, and the pulse is compressed to 15.1 fs.
Figure \ref{fig4} also indicates that the propagation in air has little influence on pulse compression.
However, the air gap is crucial to the inhibition of multi-filamentation and  keeping of vortex characteristics, where the vortex beam undergoes a self-healing process of spatial distortions \cite{Xu22}.

 \begin{figure}[htbp]
\centering
	\includegraphics[width=0.82\linewidth]
	{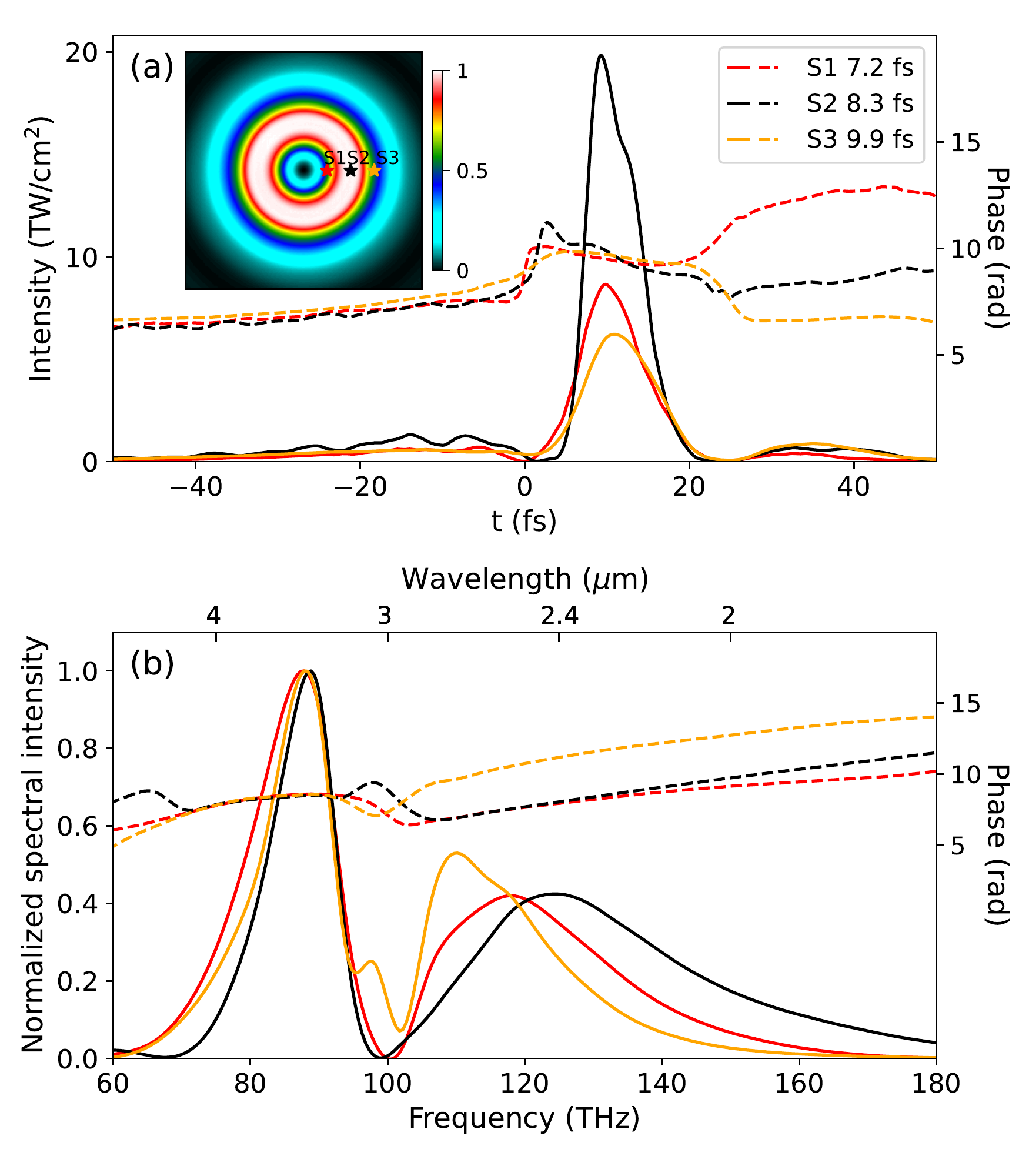}
\caption{(a) The temporal intensity and phase of three transverse positions of the laser pulse after Plate 5. (b) The corresponding spectral intensity and phase. S1, S2, S3 are marked by stars in the inset of (a), which shows the fluence distribution after Plate 5.}
\label{fig5}
\end{figure}

The influence of the spatial intensity distribution on pulse compression has also been investigated.
In Fig. \ref{fig5}(a) we show the temporal intensity and phase of different transverse positions for the laser pulse after Plate 5.
Three typical transverse positions (inner ring, intensity maximum, and outer ring) are selected, marked by stars S1, S2 and S3 in the inset.
For these three positions, the pulses are all compressed below 10 fs, and the pulse duration of S3 is larger, as we expect.
This is because the periphery has a lower intensity and experiences less spectrum broadening.
For the inner ring S1, it actually comes from the self-focusing part of the most intense ring, which has a shorter duration.
We can also see that the temporal phase slope of S2 is higher than S1 and S3, indicating that S2 has more blueshift spectral components due to stronger ionization.
The above explanation is further verified by the Fig. \ref{fig5}(b), where we show the corresponding spectral intensity and phase.
As we can see, due to SPM and ionization, the fundamental wavelength (3 $\mu$m) almost disappears, and there are two peaks in the red shift ($> 3\ \mu$m) and blue shift ($<3\ \mu$m) region. 
In the main peak of spectral intensity (3.24 - 3.75 $\mu$m), the three spectral phase curves are all flat, which means the chirp is fully compensated.
For the blue shift region, we still see a small positive chirp, and this chirp is less for S1, leading to a shorter pulse duration.
Since the uncompressed portion mainly locates at the periphery, spatial filtering technique may be used to obtain shorter pulses.

In conclusion, we demonstrate theoretically the generation of high power few-cycle vortex beam using the thin-plate scheme.
Due to the strong negative GVD of fused silica plates, the 3 $\mu$m femtosecond vortex beam undergoes self-compression, and the pulse is compressed from 3 mJ/90 fs to 2.75 mJ/15.1 fs, corresponding to about 1.5 optical cycles with a peak power of 0.18 TW. 
At the same time, the thin-plate scheme prevents destructive multi-filamentation, so that the few-cycle pulse preserves the vortex characteristics.
As a result, few-cycle vortex beam is obtained, and the conversion efficiency is as high as 91.5\% due to weak ionization.
This approach will benefit the generation of isolated attosecond vortices, opening a new perspective in ultrafast science.

\section*{Funding}
National Natural Science Foundation of China (11874056, 12074228, 11774038); Natural Science Foundation of Shandong Province (ZR2021MA023); Taishan  Scholar Project of Shandong Province (tsqn201812043); Innovation Group of Jinan (2020GXRC039).
\section*{Disclosures}
The authors declare no conflicts of interest.
\section*{Data availability} Data underlying the results presented in this paper may be obtained from the authors upon reasonable request.

\bibliography{reference}

\end{document}